%% file: main.tex
\documentclass[letterpaper]{article} 
\usepackage{aaai2026}  
\usepackage{times}  
\usepackage{helvet}  
\usepackage{courier}  
\usepackage[hyphens]{url}  
\usepackage{graphicx} 
\usepackage{natbib}  
\usepackage{caption} 
\usepackage{algorithm}
\usepackage{algorithmic}

\usepackage{xcolor}
\newcommand{\answerYes}[1]{\textcolor{blue}{#1}} 
\newcommand{\answerNo}[1]{\textcolor{teal}{#1}} 
\newcommand{\answerNA}[1]{\textcolor{gray}{#1}}

\usepackage{newfloat}
\usepackage{listings}
\usepackage{graphicx,multirow,xspace,longtable}
\usepackage{tabularx,booktabs,blindtext,url}
\usepackage{makecell,amsmath,rotating}

\usepackage[utf8]{inputenc}
\DeclareCaptionStyle{ruled}{labelfont=normalfont,labelsep=colon,strut=off} 
\floatstyle{ruled}
\newfloat{listing}{tb}{lst}{}
\floatname{listing}{Listing}
\usepackage[capitalize,noabbrev]{cleveref}
\usepackage{subcaption}
\newcommand\clearrow{\global\let\rowmac\relax}
\newcommand{\dataset}[1]{$\mathcal{D}_{\text{#1}}$}

\crefformat{section}{\S#2#1#3}
\crefformat{subsection}{\S#2#1#3}
\crefformat{subsubsection}{\S#2#1#3}

\newcommand{\para}[1]{{\vspace{.05in} \bf \noindent #1 }}
\newcommand{\parait}[1]{{\vspace{.05in} \em \noindent #1 }}

\newcommand{\eg}{e.g.,\ }
\newcommand{\etal}{et al.\xspace}

\title{Turning Trust to Transactions:\\Tracking Affiliate Marketing and FTC Compliance in YouTube's Influencer Economy}

\author {
    Chen Sun\textsuperscript{\rm 1},
    Yash Vekaria\textsuperscript{\rm 2},
    Zubair Shafiq\textsuperscript{\rm 2},
    Rishab Nithyanand\textsuperscript{\rm 1}
}
\affiliations {
    \textsuperscript{\rm 1}University of Iowa\\
    \textsuperscript{\rm 2}UC Davis\\
    \{chen-sun, rishab-nithyanand\}@uiowa.edu, \{yvekaria, zubair\}@ucdavis.edu
}

\begin{document}
\maketitle

\input{abstract}
\input{introduction}

\input{dataset_method}
\input{detecting_disclosure}

\input{responsible}
\input{related}

\input{conclusion}

\bibliography{paper,affiliates_paper,aaai2026}
\input{ethics_checklist}

\input{appendix}

\end{document}

%% file: abstract.tex
\begin{abstract}

YouTube has evolved into a powerful platform where creators monetize their influence through affiliate marketing, raising concerns about transparency and ethics, especially when creators fail to disclose their affiliate relationships.
Although regulatory agencies like the US Federal Trade Commission (FTC) have issued guidelines to address these issues, non-compliance and consumer harm persist, and the extent of these problems remains unclear.
In this paper, we introduce tools, developed with insights from recent advances in Web measurement and NLP research, to examine the state of the affiliate marketing ecosystem on YouTube. We apply these tools to a 10-year dataset of 2 million videos from nearly 540,000 creators, analyzing the prevalence of affiliate marketing on YouTube and the rates of non-compliant behavior. 
Our findings reveal that affiliate links are widespread, yet disclosure compliance remains low, with most videos failing to meet FTC standards. 
Furthermore, we analyze the effects of different stakeholders in improving disclosure behavior. 
Our study suggests that the platform is highly associated with improved compliance through standardized disclosure features.
We recommend that regulators and affiliate partners collaborate with platforms to enhance transparency, accountability, and trust in the influencer economy.
\end{abstract}

%% file: introduction.tex
\section{Introduction}
\label{sec:introduction}

YouTube has transformed into a cultural and economic force, with millions of viewers regularly turning to their favorite creators for advice, recommendations, and information.
A recent YouTube Trends survey \cite{youtube-survey-2024} found that 65\% of online users aged 14-24 identified as ``creators'', while 85\% considered themselves fans of other creators. Over half reported either spending money on their fandom or monetizing their own following. 
This shift highlights how creators influence consumer behavior by bridging entertainment and commerce.
Brands and advertisers have taken notice, increasingly investing in influencer marketing \cite{adage-2024}. A recent report projects the influencer marketing industry will reach \$24B by the end of 2024 \cite{influencermarketinghub}. 
These investments underscore creators' growing influence over consumer decisions.

Despite their strong cultural and economic influence, creators often face challenges with YouTube's frequently changing monetization policies, definitions of “advertiser-friendly content”, and unpredictable demonetization algorithms \cite{dunna_paying_2022}.
Prior research \cite{caplan_tiered_2020, hua_characterizing_2022} indicate that this instability drives creators to pursue more reliable income sources to sustain their online presence, with affiliate marketing emerging as one increasingly popular alternative \cite{syrdal2023influencer, youtube-affiliate-marketing-stats-2024}.  
Affiliate marketing is a commission-based strategy in which creators earn revenue by promoting products through personalized hyperlinks known as affiliate links. When a viewer clicks on such a link and completes an action (e.g., making a purchase), the creator receives a pre-negotiated commission.
This model offers creators financial stability by allowing them to monetize their audiences without depending solely on YouTube’s policies and algorithms.
However, this revenue model also brings new challenges around transparency and ethics, as audiences may perceive influencers as more authentic than traditional experts \cite{coco2020sponsored}.
When creators promote products without clearly indicating affiliate relationships, it can lead audiences to view recommendations as unbiased \cite{vox2022tiktok}, which may undermine trust and, in some cases, result in consumer harm.

Consumer protection agencies have long recognized the risks associated with undisclosed influencer marketing.
As early as 2009, the Federal Trade Commission (FTC) issued guidelines for endorsements in online advertising, which were updated in 2015 and again in 2023 \cite{ftc2009guides, ftcguide2023updates}.
It requires any connection between an endorser and an advertiser must be disclosed in a manner that is \textit{Clear and Conspicuous} \cite{ftcguide2023updates}.
Despite these proactive regulatory efforts, undisclosed affiliate marketing continues to be a concern.
For instance, several high-profile influencers have recently promoted cryptocurrencies and NFTs without disclosing their personal ties to these projects, costing followers millions when these ventures failed \cite{yaffe2022influencers, paul2023decrypt}.
As a result, questions remain about the overall effectiveness of these regulatory measures in safeguarding consumer interests.

Affiliate marketing is a growing monetization strategy for content creators, but little is known about its details.
How prevalent is affiliate marketing on YouTube? 
Do influencers commonly disclose their affiliate relationships, or do undisclosed affiliate links persist across the platform?
When disclosures are made, do they align with the standards set by the FTC?
And perhaps more importantly, which stakeholders appear most closely associated with improved disclosure compliance?
A key challenge in addressing these questions is the complexity of identifying affiliate links and disclosures at scale. Our work tackles these challenges by developing new methods, leveraging recent advances in Web measurement and NLP, to accurately identify textual affiliate links and related disclosures (see \Cref{sec:methods} for details).
While affiliate promotions and disclosures can appear in spoken audio or on-screen visuals during the video, our study focuses on written affiliate links in video descriptions and the YouTube Shopping shelf and their accompanying textual disclosures, in line with FTC guidance requiring that disclosures match the format and placement of the promotional content \cite{ftcguide2023updates, FTC_frequently_ask_questions}.
Using these tools, we analyze over 2 million YouTube videos from nearly 540,000 unique creators across a diverse range of content, providing a comprehensive view of affiliate marketing on YouTube.
In particular, we address the following questions.

\begin{itemize}
    \item {\em RQ1. How prevalent is affiliate marketing on YouTube, and to what extent do creators disclose their affiliate relationships in compliance with FTC guidelines? (\cref{sec:detecting_disclosure})}  
    We introduce a new approach for identifying affiliate links, addressing limitations of prior research to measure the prevalence of affiliate marketing across our dataset. Then we deploy a series of transformer-based classifiers to detect affiliate marketing disclosures. 
    This combined approach allows us to quantify the prevalence of affiliate marketing across the platform and assess whether creators comply with the FTC’s ``clear and conspicuous'' standard.
    We analyze patterns by content category and channel size, revealing that affiliate marketing is widespread, particularly in product-oriented communities and larger creators, however, compliance with FTC standards remains low.

    \item {\em RQ2. How do different stakeholders influence affiliate disclosure compliance on Youtube?} (\Cref{sec:responsible})
    Building on our discovery that most affiliate containing videos fail to meet FTC disclosure standards, we investigate the roles that three different stakeholders (YouTube, regulators, and affiliate partners) play in shaping compliance outcomes.
    Our analysis shows that videos linked to affiliate partners that provide public disclosure guidelines, as well as those published after FTC policy updates, exhibit modest improvements in compliance. In contrast, videos featuring platform-level mechanisms, such as the auto disclosure tool in the YouTube Shopping tab, show substantially higher compliance rates. While these findings are observational and do not imply causality, they highlight the potential for platforms to serve a central role in providing standardized disclosure solutions.

\end{itemize}

Together, these research questions address key gaps in our understanding of affiliate marketing and disclosure practices within YouTube's influencer economy.  
By examining both the prevalence of affiliate links and compliance with FTC disclosure standards, this study offers a comprehensive view of the ongoing consumer protection and trust challenges within the affiliate marketing ecosystem. 
Additionally, by analyzing how compliance varies across different stakeholder contexts, our findings suggest that platform-integrated disclosure tools are highly associated with improved transparency and compliance. Regulators and affiliate partners should collaborate with platforms to help design such tools and support monitoring and oversight.

%% file: dataset_method.tex
\section{Data and Methods} \label{sec:methods}

\begin{figure*}[!t]
    \centering
    \includegraphics[width=1\textwidth]{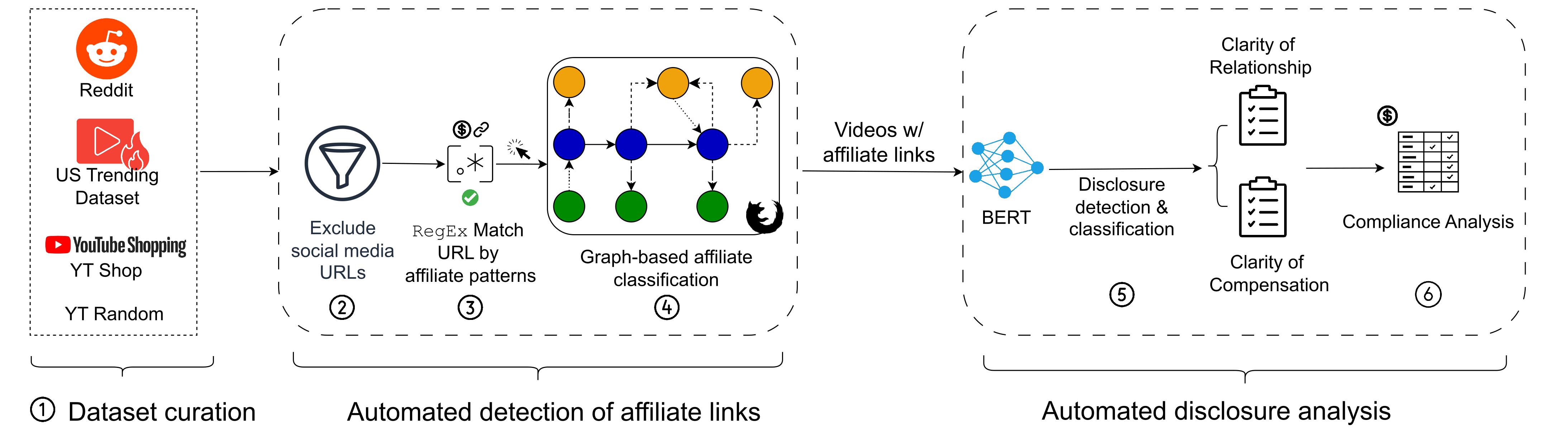}
    \caption{Overview of our approach for detecting affiliate links and analyzing disclosure compliance.}
    \label{fig:study_overview}
\end{figure*}

Our study on identifying affiliate links and related disclosures consists of three main components: video collection (\Cref{sec:methods:data}), affiliate link detection (\Cref{sec:methods:links}), and disclosure classification (\Cref{sec:methods:disclosures}).
An overview of the full analysis pipeline is illustrated in \Cref{fig:study_overview} \footnote{Code and analysis results are available at: \url{https://osf.io/3xpwg/?view_only=fd0cb236942142ad915d9d0c1534a8e0}}.

\begin{table}[t]
\centering
{
\small
\begin{tabular}{l c c c}
\toprule
{\bf Source} & {\bf \# videos} & {\bf \# hyperlinks}  & {\bf \# channels}  \\
\midrule
Reddit ($\mathcal{D}_{\text{Reddit}}$)              & 1.88M & 3.78M      & 480.0K  \\
Random ($\mathcal{D}_{\text{Random}}$)              & 64.2K & 216.0K      & 58.7K \\
Trending ($\mathcal{D}_{\text{Trending}}$)  & 46.6K & 120.1K     & 7.5K  \\
Shopping ($\mathcal{D}_{\text{Shopping}}$)  & 10.9K  & 90.2K      & 2.9K   \\
\midrule
\bf{Total (\dataset{})} & 2.00M & 4.13M & 539.9K \\
\bottomrule
\end{tabular}
}
\caption{Summary of our dataset. Numbers reflect the number of \textit{unique} videos, hyperlinks, and channels per source.}
\label{tab:methods:data}
\end{table}

\subsection{YouTube Video Collection} \label{sec:methods:data}
Our goal was to create a sample of videos that could simultaneously achieve three objectives: represent  
(1) the videos consumed by the general YouTube audience across a wide range of topics; 
(2) the behaviors within the affiliate marketing ecosystem before and after 2018, a pivotal year marked by increased FTC enforcement actions for endorsement guideline violations and the repercussions of the ``Adpocalypse''; and  
(3) the videos chosen through different YouTube promotion mechanisms.
To achieve this, we collected videos from four distinct sources. The Reddit dataset (\dataset{Reddit}) includes YouTube links extracted from Reddit posts, and the Random dataset (\dataset{Random}) was generated using a five-character prefix method to randomly sample video IDs \cite{mathur2018endorsements}. To capture algorithmically promoted content, we included the YouTube Trending dataset (\dataset{Trending}), compiled from a Kaggle source and continuous API crawling of YouTube’s US and global Trending tabs every 15 minutes. Lastly, the Shopping dataset (\dataset{Shopping}) consists of influencer-promoted product videos, scraped every 15 minutes from YouTube’s Shopping tab, which features content selected by YouTube’s shopping program. 
\Cref{tab:methods:data} summarizes the contributions of each source. Overall, our dataset contains 2 million unique publicly listed videos, uploaded between 01/2015 and 12/2024, spanning 539.9K channels and 16 video categories.
For each video, we extract all hyperlinks from two sources: the video description and the tagged products section in YouTube’s shopping icon, if present. In total, our dataset contains 4.13M unique hyperlinks.

\subsection{Detecting Affiliate Links} \label{sec:methods:links}
Our next goal was to identify textual affiliate links in each video, including those found in the video description and YouTube’s shopping shelf. 
Prior approaches~\cite{mathur2018endorsements, hua_characterizing_2022, swart_is_2020, ballard_conspiracy_2022} relied on regular expressions or word-domain co-occurrences; however, such methods can be easily evaded when link structures change or when dual-purpose and shortened domains are used.
To address this, we design a scalable, redirect-chain-based method that captures browser behaviors triggered by clicking each link, making our approach effective and robust against these challenges.
At a high level, our approach to detecting affiliate links is based on the insight that, due to the mechanisms underlying affiliate linking, the characteristics of the resulting redirects and browser behaviors differ fundamentally from those of non-affiliate links.  
This approach was inspired by recent research from Munir \etal \shortcite{munir2024purl}, who used a similar method to identify and block tracking-related data exfiltration from a user's browser via URL. 
We operationalize this insight through the three-phase approach described below.

\para{\bf Phase 1: Labeling known hyperlinks.}
The first phase involved applying the regular expression-based approaches used by Mathur \etal \cite{mathur2018endorsements} and Swart \etal \cite{swart_is_2020} to identify non-shortened affiliate links with well-known structures.
For instance, non-shortened links to Amazon affiliate storefronts follow the format {\tt www.amazon.com/shop/<store name>}, while non-shortened links to affiliate-promoted products include the {\tt tag=<affiliate id>} key-value pair in the URL.  
We manually crafted a set of regular expressions using a sample of the top 1000 most frequently observed domains within our dataset. The URLs and public information associated with each of these domains were manually examined to identify patterns commonly associated with affiliate programs (e.g., containing an {\tt affiliate\_id=<id>} key-value pair). Whenever possible, we created regular expressions based on these patterns, and these regular expressions were then used to automatically label any matching links as affiliate links. 
Since they were very common, we also crafted regular expressions to identify links to creators' profiles on popular social media platforms (YouTube, Facebook, Instagram, Twitter, Snapchat, and TikTok), which were automatically labeled as non-affiliate links.  In total, 29\% of the hyperlinks in our dataset were labeled using this approach. 

\para {\bf Phase 2: Generating interaction graphs.}
At the end of Phase 1, 71\% of the hyperlinks could not be labeled as affiliate or non-affiliate links. Next, we applied an approach similar to that of Munir \etal \cite{munir2024purl}, instrumenting the browser to create an interaction graph that captures the browser’s behavior from the moment a hyperlink is clicked until it reaches a final landing page (i.e., one with no further redirects).  
We loaded each video in our dataset using OpenWPM (v0.17.0) \cite{englehardt2016online}, wrapped around Firefox (v102) and driven by Selenium (v4.13), while following best practices for interaction-driven crawling \cite{ahmad_apophanies_2020}.  
Each hyperlink in the video description was then clicked to generate an interaction graph which captures the resulting sequence of redirects and information flows between the user's browser and other entities. More details on these graphs are provided in \ref{sec:appendix:affiliate_link_detection}

\para {\bf Phase 3: Feature extraction and classification.}
We then extracted features from each interaction graph in our dataset, focusing on two types of features: graph structural features and redirect features. Structural features captured the general structure and connectivity of the interaction graph (e.g., graph density, centrality, average shortest path between all pairs of nodes), while redirect features focused on the characteristics of the redirects (e.g., number of query parameters attached to network nodes, length of the redirect chain, Shannon entropy of key-value pairs).
To create a ground truth dataset for classifier development, we used features extracted from the interaction graphs of the affiliate links identified in Phase 1 as our labeled `Affiliate' samples.  
Two authors then manually inspected the most frequently linked landing page domains within YouTube descriptions, along with the domains within their redirect chains, until they identified 1,000 landing page domains without any affiliate-link indicators.  
The annotating authors shared 100 randomly selected overlapping domains, which they each evaluated independently, discussing challenging cases halfway through. In the first half of their non-affiliate annotations, they achieved a Cohen's $\kappa$ of 77\%. Following discussion, Cohen's $\kappa$ increased to 85\% for the remaining shared domains. This process is described in more detail in \ref{sec:appendix:affiliate_link_detection}
The features extracted from links associated with these 1,000 domains served as our `Non-affiliate' samples.
We then used these samples to train a random forest classifier due to its robustness to overfitting and success in prior interaction graph-based classification tasks~\cite{iqbal2020adgraph, munir2024purl}.
We split the ground truth dataset into three parts: a training and testing set (60\%) and two holdout sets (20\% each), one holdout set had domains seen during training while the other only contained entirely unseen domains to assess the generalizability of the model.
Using the training/testing set, we applied 5-fold cross-validation and performed a grid search to optimize hyperparameters. 
Our affiliate link classifier achieved strong performance, with an F1-score of 92.8\%–96.1\% across the two holdout sets. In contrast, prior regex-based approaches achieved F1-scores of 18.5-23.1\%. This dramatic difference in performance is largely due to the high prevalence of shortened URLs and affiliate-related information flows that are only visible through analysis of redirect chains.

\subsection{Detecting Affiliate-Related Disclosures} \label{sec:methods:disclosures}
Our final goal was to evaluate whether affiliate disclosures are present and whether they meet FTC standards.  
We focused exclusively on English-language text disclosures in the description box and the shopping tab, and did not analyze audio or on-screen visual disclosures.

\para{Operationalizing FTC standards.} 
We carefully reviewed the FTC’s updated Endorsement Guides~\cite{ftcguide2023updates}, accompanying FAQs~\shortcite{FTC_frequently_ask_questions}, public comments~\shortcite{FTC_responses_to_public_comments}, and enforcement cases~\shortcite{ftc2020_warning_Teami_case_influencers, ftc2023_warning_amer_bev}. Across these documents, we found that the FTC consistently evaluates disclosure compliance based on whether a disclosure is ``clear and conspicuous''.  
In its 2023 update, the FTC further clarified this standard as: ``a disclosure is difficult to miss (i.e., easily noticeable) and easily understandable by ordinary consumers.''~\shortcite{ftcguide2023updates}.
Based on these insights, we decomposed this standard into two dimensions. (1) \textit{Clarity of Compensation} which assesses whether the disclosure use plain, understandable language to explain that the creator receives financial compensation from the associated link, aligning with the emphasis on `clear'; and (2) \textit{Clarity of Relationship} which evaluates whether the disclosure can be reasonably attributed to the affiliate links that makes it ``difficult to miss'', addressing the requirement for `conspicuous'.
To capture disclosure quality, we defined three labels for each dimension informed by FTC guidance. For the ``Clarity of Compensation'' dimension, we introduce: (1) \textit{Clear}: Disclosures that explicitly state a compensatory relationship (e.g., “I earn a small commission”); (2) \textit{Ambiguous}: Vague or implied support language (e.g., “Support me by clicking below”); and (3) \textit{None}: Mentions of affiliate terms without explaining compensation (e.g., “affiliate link”), which the FTC considers as non-compliant.
For the ``Clarity of Relationship'' dimension, we define: (1) \textit{Explicit}: Each affiliate link is individually disclosed; (2) \textit{Grouped}: A single disclosure applies to a list of affiliate links; and (3) \textit{Mixed group}: A vague statement that covers affiliate and non-affiliate links.

\para{Disclosure annotation.} 
Once we translated FTC guidelines into computable labels, we annotated textual content from the description box and shopping tab (from \dataset{Shopping}).
For the videos from \dataset{Shopping}, each affiliate link is clearly tagged by YouTube with “Earns commission.” We label these as clearly compliant disclosures for both compensation and relationship dimensions.
For affiliate links appearing in video descriptions, we developed classifiers to assess the presence and quality of disclosures.
We first pre-processed video descriptions into sentence-level segments. Next, we manually annotated 2K affiliate video sentence-level segments, selected randomly and equally across all four datasets.
Each sentence was initially annotated as a `Disclosure' or `Non-disclosure'. If multiple contiguous sentences are identified as disclosures, we merged them into a single disclosure segment to preserve contextual meaning.
For segments annotated as disclosures, we manually tagged their ``Clarity of Compensation'' and  ``Clarity of Relationship'' using the labels defined above.
To ensure annotation quality, two annotators independently labeled an overlapping 10\% of the sample. The inter-annotator agreement was high, with a Cohen’s $\kappa$ of 0.86, indicating strong reliability.

\para{Classifier development.} 
Using the annotated dataset, we trained three BERT-based classifiers, where the \textit{input} is a sentence-level text segment and the \textit{output} indicates whether the sentence is part of a disclosure and, if applicable, the ``Clarity of Compensation'' and the ``Clarity of Relationship''.
For instance, given the input “If you click on this link and purchase a product, I get a small commission at no cost to you,” the model outputs “Disclosure” for presence, “Clear” for compensation, and “Explicit” for relationship.  
We used the BERT model \cite{devlin2019bert} for encoding text data and the \texttt{BertForSequenceClassification} model for our classification tasks. This model adds a classification layer over the BERT architecture, producing logits for each class based on a given input.  
To prevent overfitting, we balance the training set via undersampling and apply early stopping if the F1 score does not improve for three consecutive epochs.
20\% of our ground truth annotations were saved as a holdout set for evaluation of each underlying classifier. On this set,  the models achieved high F1-scores: 96.8\% for disclosure detection, 98.6\% for clarity of compensation, and 97.0\% for clarity of relationship.
For comparison, we also implemented a keyword detector using common disclosure markers (e.g., \#ad, sponsored, affiliate link) based on prior work~\cite{bertaglia2025influencer, bertaglia2024monetisation} and FTC’s Endorsement Guides~\cite{FTC_frequently_ask_questions}. 
With this broad disclosure keyword list, we achieved an F1-score of 86.4\%.
However, these keyword-based methods cannot assess the clarity of disclosures because of their inability to capture semantic and positional relationships. 
Overall, our classifiers not only achieve higher accuracy but also assess disclosure compliance and clarity dimensions in alignment with FTC standards.
We provide detailed annotation codebook with label definitions, classification performance metrics, and error analysis in~\ref{sec:appendix:affiliate_disclosure_classification}

\subsection{Ethical Considerations}
Our study did not involve any direct interactions with YouTube creators or users.
For data collection, we relied on the YouTube API whenever possible (e.g., for collecting Trending videos and metadata for all videos in our dataset).  
We only used an automated crawler to click hyperlinks shown in video descriptions, which is a necessary step for generating the interaction graphs that underpin our affiliate link detection. 
To minimize the impact on content creators and the YouTube recommendation system, each video is visited only once, and each unique link is clicked only once. Our crawler ignored duplicate URLs and common social media or profile links.  
Overall, this automated crawl adhered to YouTube’s implicit rate limits to avoid being classified as “suspicious traffic” and was conducted exclusively for this study between 04/2024 and 12/2024.
The information we recorded included video metadata, description text, URLs, and browser interactions. In our public release, we include only publicly available metadata and links. All raw interaction logs (e.g., cookies and network traces) were used solely for model development.

%% file: detecting_disclosure.tex
\section{Prevalence of Affiliate Marketing and Quality of Disclosures} \label{sec:detecting_disclosure}

We now answer the question \textit{RQ1: How prevalent is affiliate marketing on YouTube, and to what extent do creators disclose their affiliate relationships in compliance with FTC guidelines?}
We first examine overall affiliate link usage, disclosure presence and clarity, and key reasons for non-compliance. We then explore how affiliate and disclosure practices vary across content categories and channel sizes. 
We introduce seven metrics. Four capture the prevalence of affiliate marketing:  fraction of videos that contain affiliate links (AV); fraction of channels that use affiliate links in their videos (AC);  number of affiliate links per video (NALPV); and fraction of links that are affiliate links per video(FLAL). 
Three assess the quality of affiliate disclosures: fraction of videos that clearly compliant with FTC standards (CC);  fraction of videos that are possibly compliant, where disclosures are present but exhibit ambiguity in either compensation or relationship clarity (PC); and fraction of videos that are clearly non-compliant (NC).

\begin{table}[t]
    \centering
\resizebox{\columnwidth}{!}{%
    \begin{tabular}{lllc}
    \toprule
         {\bf Compliance}   &  {\bf Clarity of }    & {\bf Clarity of}      & {\bf Fraction of} \\
         {\bf status}       &  {\bf compensation }  & {\bf relationship}    & {\bf videos (\%)} \\
         \midrule
         CC         &   Clear               &   Explicit or Grouped &   12.20\\
         \midrule
         \multirow[c]{2}{*}{}
    
                            &   Clear               &   Mixed Group         &   9.31  \\
        \multirow[c]{2}{*}{PC}
                            &   Ambiguous           &   Explicit or Grouped &   2.95  \\
                            &   Ambiguous           &   Mixed Group         &   6.35  \\
         \cline{2-4}
                            &  \multicolumn{2}{c}{{\bf Total}}              &   18.61 \\
         \midrule
         \multirow[c]{4}{*}{NC}                            
                            &   Absent              &   Explicit or Grouped &   12.32 \\
                            &   Absent              &   Mixed Group         &   2.68  \\
                            &   Absent              &   Absent              &   54.19 \\
        \cline{2-4}
                            &  \multicolumn{2}{c}{{\bf Total}}              &   69.19 \\
        \bottomrule
    \end{tabular}
}    
    \caption{Breakdown of disclosure clarity in \dataset{Affiliate}. Definitions of CC (clearly compliant), PC (possibly compliant), and NC (non-compliant) disclosures are in \Cref{sec:methods:disclosures}.}
    \label{tab:rq2:reasons}
\end{table}

\para{Affiliate links are common, but disclosures are rare.}  
Using the affiliate link and disclosure classifiers introduced in \Cref{sec:methods}, we analyzed the hyperlinks and texts associated with over 2M videos in our dataset \dataset{}.
We identified 352.7K unique affiliate links across 146.8K videos (7.35\% of all videos in \dataset{}) and 36.8K channels (6.81\% of all channels in \dataset{}), representing 8.55\% of all hyperlinks in the dataset. We refer to these as the \dataset{Affiliate}.
On average, videos in \dataset{Affiliate} contained 5.28 unique affiliate links, accounting for 38.02\% of all links in their descriptions.
At the channel level, the average channel used 5.20 affiliate links in their videos, representing 46.09\% of all hyperlinks in their videos content.  
Despite this widespread use, disclosure remains rare. Among 139.9K English-language videos in \dataset{Affiliate}, only 45.52\% videos from 47.17\% channels included any form of disclosure. And just 12.20\% of videos and 12.98\% of channels provided disclosures that clearly met FTC guidelines, demonstrating both compensation and relationship clarity.  
\Cref{tab:rq2:reasons} provides a detailed breakdown of affiliate disclosures by compliance status. We consider videos clearly compliant with FTC if they are labeled as clear in compensation and explicit or grouped in relationship clarity.
Those with ambiguous or mixed-group disclosures are labeled as possibly compliant, while disclosures that are entirely missing or fail to describe any compensation to the creator are classified as non-compliant.  
Notably, 69.19\% of affiliate videos were non-compliant, with 54.19\% lacking any disclosure at all, indicating that most creators neglect to inform viewers about their financial relationships with affiliate partners.  
Among the 18.61\% labeled as possibly compliant, the most common issue is the disclosure do not clarify which of the links in their descriptions are affiliate links (e.g., “some of the links below are affiliate links”).  
These findings suggest that while affiliate marketing is pervasive on YouTube, FTC compliance remain persistently low from 2015–2024, raising serious concerns about transparency and consumer trust.

\para{Affiliate marketing has grown since the Adpocalypse, but improvements in disclosures remain modest, especially among larger channels.}
\Cref{tab:rq1:popularity_stacked} compares affiliate marketing and disclosure practices across videos uploaded before and after 2018, revealing several notable trends. Overall, affiliate marketing has grown substantially across all channel size since the Adpocalypse. Between 2015–2018 and 2018–2024, the percentage of videos and channels containing affiliate links increased from 5.3\% to 8.5\% and from 5.2\% to 7.7\%, respectively. 
This growth is especially pronounced among smaller channels ($<$100K subscribers), which nearly doubled their affiliate video rate from 3.22\% to 6.38\%, and among mid-sized channels (100K–1M), which showed the largest overall increase in affiliate prevalence from 8.66\% to 12.73\%.
Further, we found that large channels ($>$1M subscribers) exhibited highest affiliate usage across both time periods.
While smaller creators used affiliate marketing less frequently overall, when they did, their videos contained a higher fraction of affiliate links, reflecting a stronger emphasis on monetizing each opportunity.
Focusing on disclosure practices, the fraction of affiliate videos with clearly compliant disclosures more than doubled, increasing from 5.5\% to 14.5\% across all channel sizes. 
Improvements were notable among small and mid-sized channels where non-compliance dropped by 17.6\% and 17.3\%, respectively. In contrast, large channels (1M+ subscribers) remained the least compliant, only decreasing by 9.3\%. 
The negative correlation between channel size and compliance was confirmed by a statistically significant Pearson correlation ($\rho = -.05$, $p < .05$). These findings align with prior research suggesting that popular influencers may avoid disclosure due to fear of audience backlash or perceived damage to authenticity \cite{cheng2024reputation, berger2016research}. 
While these results indicate some progress in disclosure practices, overall compliance remains low. In particular, larger creators, who have the widest influence, remain the most lacking in disclosure compliance.

\begin{table}[t]
\centering
\resizebox{\columnwidth}{!}{%
\begin{tabular}{lcccc}
\toprule
 & 1--100K & 100K--1M & 1M+ & Total \\
\midrule
AV (\%) & {6.4} (+3.2) & {12.7} (+4.0) & \textbf{11.4} (+3.4) & 8.5 (+3.2) \\
AC (\%) & \textbf{6.0} (+2.3) & \textbf{17.0} (+3.0) & \textbf{20.5} (+3.4) & 7.7 (+2.5) \\
NALPV & {5.8} (+1.7) & {6.7} (+2.6) & \textbf{4.6} (+1.1) & 5.7 (+1.7) \\
FLAL (\%) & \textbf{46.0} (+1.8) & \textbf{38.5} (+7.3) & \textbf{29.2} (+4.9) & 39.5 (+5.6) \\
NC (\%) & {61.4} (-17.6) & {62.3} (-17.3) & \textbf{74.7} (-9.3) & 65.1 (-15.8) \\
PC (\%) & \textbf{23.8} (+7.9) & \textbf{20.7} (+6.1) & \textbf{14.5} (+4.2) & 20.3 (+6.7) \\
CC (\%) & \textbf{14.8} (+9.7) & {17.1} (+11.3) & {10.8} (+5.1) & 14.5 (+9.0) \\
\bottomrule
\end{tabular}
}
\caption{Prevalence of affiliate links and rates of (non-) compliance by channel size for videos uploaded between 2018--2024. Values in parentheses show changes from 2015–2018. 
Bolded cells indicate values that differ significantly from other channel size tiers within the same 2018--2024 period ($p < .05$). 
Comparing across time periods for each channel size, all differences were found to be statistically significant ($p < .05$) and thus not highlighted.
A two-sided $z$-test of proportions was used to test the significance of differences in percentage of affiliate videos and channels, as well as the compliance rate across sources. A two-sided Welch's $t$-test was used to test the significance of differences in number of affiliate links and affiliate fraction of links per video.}
\label{tab:rq1:popularity_stacked}
\end{table}

\begin{table}[t]
\centering
{
\small
\begin{tabular}{lccccc}
\toprule
\textbf{Category}  & \textbf{AV} & \textbf{FLAL} & \textbf{NC} & \textbf{PC}& \textbf{CC} \\

 & \textbf{(\%)} & \textbf{(\%)} & \textbf{(\%)}  & \textbf{(\%)}  & \textbf{(\%)}  \\
\midrule
Autos \& Veh  & 17.1  & \bf 52.2 & 65.3 & 20.0 & 14.7\\
Comedy & 5.8 & 29.3  & \bf 92.3 & 4.9 & 2.9  \\
Education & 14.6  & 36.1  & 66.1 & 21.3 & 12.6\\
Entertainment & 7.3  & 35.2 & 75.1 & 12.9 & 12.0 \\
Film \& Anim & 7.0 & 34.9 & 78.0 & 12.8 & 9.2  \\
Gaming & 5.3 & 28.4 & 75.9 & 19.2 & 4.9 \\
Howto \& Style & \bf 26.6  & \bf 49.4 & 55.9 &\bf 24.6 & \bf 19.5 \\
Music & 3.4  & 33.5  & 81.6 & 12.4 & 6.0  \\
News \& Politics & 3.2 & 34.2  & 79.1 & 17.3 & 3.6  \\
Nonprofits & 1.7  & 38.2 & 70.2 & 10.1 & \bf19.6 \\
People \& Blogs & 7.7 &  \bf52.8 & 65.4 & 14.7 & \bf19.9 \\
Pets \& Animals & 9.9 & \bf54.8  & 54.5 & \bf 26.1 & \bf19.4 \\
Sci \& Tech & \bf 23.8 &  41.7 & 54.4 & \bf 25.3 & \bf 20.3  \\
Sports & 4.9 & 36.3 & 76.3 & 14.3 & 9.4 \\
Travel \& Events & 11.8 & 47.0  & 68.9 & 17.9 & 13.2 \\
\bottomrule
\end{tabular}
}
\caption{Prevalence of affiliate links and rates of (non-) compliance of videos in \dataset{Affiliate} across top 15 categories. Bolded cells indicate that the value was a full standard deviation above the mean for the corresponding metric computed over all subgroups.
}
\label{tab:rq2:category}
\end{table}

\para{Product-oriented communities use and disclose affiliates more.}  
\Cref{tab:rq2:category} presents affiliate marketing prevalence and compliance rates across the top 15 video categories in \dataset{Affiliate}. Overall, the prevalence of affiliate links varied across categories, but disclosures consistently lacked compliance with no category achieving a clear compliance rate $>$ 20.3\%.  
Categories such as \textit{Howto \& Style} and \textit{Science \& Technology} showed higher affiliate activity, with their affiliate-related metrics exceeding the mean by at least one standard deviation. 
A review of a random sample ($n$ = 50) of affiliate videos in these categories revealed a focus on product reviews and recommendations (\eg makeup and gadget reviews). 
Interestingly, they were also found to have statistically higher disclosure rates than categories with lower affiliate link prevalence (\textit{z-}test of proportions; \textit{p} $<$ .05), suggesting that these creators may be more aware of regulatory disclosure requirements. However, they also ranked highest in ``Possibly Compliant'' metric, suggesting that these creators are attempting to disclose affiliate relationships, but do so with statements that lack clarity and precision.
Conversely, categories such as \textit{Music}, \textit{Comedy}, and \textit{Gaming} were found to have the lowest affiliate prevalence (AV $<$10\%) and the lowest clear compliance rates (CC $<$5\%).
Overall, these patterns suggest that: (1) affiliate marketing economics are especially well-suited for communities focused on consumer products, where product recommendations drive engagement and conversions; and (2) although product-focused categories are more compliant, there remains a large compliance gap across all content categories either due to unawareness of disclosure mandates or a lack of understanding of their requirements.

\para{Takeaways.} 
Affiliate marketing is widespread on YouTube with AV and AC rising from 5.3\% and 5.2\% before 2018 to 8.5\% and 7.7\% after 2018.
This increase, most pronounced among smaller channels, aligns with the Adpocalypse, monetization instability, and expanded access to influencer marketing.  
Despite this widespread adoption, our analysis reveals an abysmal state of compliance with disclosure requirements within the affiliate marketing ecosystem with over 65\% of all videos being non-compliant, exacerbated by the practices of larger channels.
In parallel, we observed that product-focused categories rely more heavily on affiliate marketing through product reviews and that these creators were more likely to disclose their affiliate relationships.

%% file: responsible.tex
\section{Stakeholder Influence on Compliance}
\label{sec:responsible}
Given the widespread use of affiliate marketing and the lack of FTC-compliant disclosures on YouTube, we now turn to our second question: 
\textit{RQ2. How do different stakeholders influence affiliate disclosure compliance on Youtube?}
In this section, we examine three key stakeholders: regulators, affiliate partners, and the platform (YouTube), by analyzing their respective roles in improving compliance.

\para{Efforts by regulators show limited impact on disclosure compliance.}
Despite FTC actions and growing public attention, our findings suggest that current regulatory efforts have limited impact on improving disclosure compliance. Disclosure rates have improved modestly since 2018, yet 65\% of affiliate videos remained non-compliant and only 14.5\% of affiliate videos clearly complied with FTC standards in the post-2018 period.
As the findings in \Cref{sec:detecting_disclosure} reveal, large channels and certain content categories, such as \textit{Comedy} and \textit{Gaming}, exhibit worse compliance performance, highlighting that regulatory effects vary across creators and content types.
Even in product-centric categories, where disclosure awareness might be expected to be higher, many disclosures remain vague or entirely absent. This high frequency suggests that interpreting and applying the FTC’s guidance remains challenging in practice.
Our own attempt to interpret the FTC requirement involved extensive research of enforcement records, legal frameworks, and policy documents -- an interpretive burden unlikely to be feasible for the average creator. This difficulty points to a broader issue: the FTC’s current disclosure requirements may lack sufficient clarity and accessibility to support large-scale adoption. This challenge is compounded by the sheer scale of the influencer ecosystem, which far exceeds the FTC’s enforcement capacity.
Collectively, our observations indicate that regulatory efforts alone have not meaningfully improved disclosure compliance at scale. Stronger interventions are needed to ensure more consistent and effective disclosure practices.

\begin{table}[t]
\centering
\resizebox{\columnwidth}{!}{%
\begin{tabular}{lccc}
\toprule
\textbf{Comparison} & \textbf{Metric (m)} & \textbf{$\Delta^{(m)}$ }  & \textbf{Effect (95\% CI)} \\
 
\midrule
\multirow[c]{3}{*}{$\mu^{(m)}_{\text{Guidance}} - \mu^{(m)}_{\text{No guidance}}$}
& NC & -3.53 & [-19.70, 13.63]\\
& PC  & -7.58 & [-22.73, 7.58]  \\
& CC & 11.11 & \textbf{[1.52, 21.21]} \\
\midrule
\multirow[c]{3}{*}{$\mu^{(m)}_{\text{Shopping}} - \mu^{(m)}_{\text{RR}}$} 
& NC & -44.78  &  \textbf{[-48.71, -40.63]}  \\
& PC & 0.13 & [-3.42, 3.57]\\
& CC & 44.65 & \textbf{[41.03, 48.32]}  \\
\bottomrule
\end{tabular}
}
\caption{Effect of affiliate program guidance (top) and platform-level built-in tools (bottom) on compliance metrics. Values in bold indicate a statistically significant effect size at the corresponding confidence level.}
\label{tab:rq2:bootstrap}
\end{table}

\para{The provision of disclosure guidelines from affiliate partners improves disclosure quality.} 
Our analysis identifies a total of 6.0K unique affiliate marketing partners, illustrating the scale of the influencer marketing industry. Among these, the Amazon Affiliate Program dominates the ecosystem, appearing in 51.8\% of all affiliate videos and 56.7\% of affiliate channels, and ranking the highest across every category and channel size. Other common partners include large networks (e.g., Impact Radius, ShareASale, Rakuten), brand-specific programs (e.g., Epidemic Sound, Apple, Tubebuddy), and general merchants like eBay. 
According to FTC guidelines, affiliate partners are expected to take reasonable steps to ensure their affiliates comply with endorsement disclosure regulations. One common approach is to provide publicly available instructions or best practices related to disclosure. 
To examine whether such guidance affects compliance, we focused on the top 30 most prevalent affiliate networks in \dataset{Affiliate}. We compared compliance rates between videos containing affiliate links to networks with disclosure guidance and those linking to networks without it.
We manually reviewed each partner’s website to assess whether it provided guidance, such as requiring affiliates to disclose their links, referencing FTC compliance, or offering example disclaimers\footnote{Full documentation for all 30 partners is available at: \url{https://osf.io/3xpwg/?view_only=fd0cb236942142ad915d9d0c1534a8e0}}.
Unfortunately, we could not locate such disclosure guidelines for 15 of these networks, while the remaining networks publicly shared such guidance.

\parait{Measuring the effect of guidance.} 
We quantified the differences in compliance rate using mean differences, defined as effect sizes. For each metric 
$m \in \{$\text{clearly compliant (CC)}, \text{possibly compliant (PC)}, \text{non-compliant (NC)} $\}$, the effect size is \( \Delta^{(m)}_{\text{Guidance} - \text{No guidance}} = \mu^{(m)}_{\text{Guidance}} - \mu^{(m)}_{\text{No guidance}} \), where \( \mu^{(m)}\) represents the average compliance rate for affiliate videos associated with partners that either provide disclosure guidance or do not.
%
To reduce bias from known confounders, we apply equal-weight sampling stratified by guidance presence, video category, channel size, and platform source. We restricted the sample to post-2018 videos to account for regulatory changes and balanced samples across affiliate partners to avoid dominance by large programs (e.g., Amazon).
To assess statistical significance, we applied bootstrap resampling with 10K iterations to estimate 95\% confidence intervals (CIs) for each effect size. Following established methodology~\cite{diciccio1996bootstrap, efron1987better}, we considered an effect significant if the 95\% CI excluded zero. 
The results, shown in the top three rows of \Cref{tab:rq2:bootstrap}, indicated that disclosure guidance was associated with a substantial and statistically significant increase in clear compliance (+11.11\%). We also observed that the presence of guidance reduced non-compliance (–3.53\%) and possible compliance (-7.58\%), although these two are not statistically significant. 
%
%
Overall, this suggests affiliate partners who offer disclosure guidance are associated with improved disclosure quality. However, the results are observational, not causal, and should be interpreted with caution. First, the analysis was limited to the top 30 affiliate programs, so the results may not generalize to the broader ecosystem. Second, we did not account for compliance measures that affiliate partners may implement behind the scenes, such as suspending commissions or terminating non-compliant accounts.

\begin{table*}[ht]
\centering
\small
{
\begin{tabular}{ccccccccc}
\toprule
 \textbf{Period} & \textbf{Source} & {\bf AV (\%)} & {\bf AC (\%)} &  {\bf NALPV} & {\bf FLAL (\%) } & \textbf{NC (\%)} & \textbf{PC (\%)}  & \textbf{CC (\%)} \\
\midrule
\multirow[c]{3}{*}{8/20-12/23} 
& \dataset{Trending}  & \bf 15.01 & \bf17.95 & 3.64 & 25.50  & \bf 86.43 & 8.53 & 5.04 \\
& \dataset{RR}  &  8.47 &  7.94 &  \bf5.71 &  \bf38.81  & 64.02 & \bf 22.36 & \bf 13.62\\
\cline{2-9}
& $\Delta_{\text{Trending - RR}}$  & 6.54 & 10.02 & -2.07 & -13.31  & 22.41 & -13.83 & -8.58 \\
\midrule
\multirow[c]{3}{*}{1/24-12/24} 
& \dataset{Shopping} & \bf50.06	 & \bf51.52 & \bf10.20 & \bf 79.05       & 20.70 & 17.39 & \bf 61.90\\
& \dataset{Trending}  & 15.81 & 18.06 & $4.19^*$ & $29.03^*$  & \textbf{83.61$^*$} & 10.15 & 6.24 \\
& \dataset{RR}  & $6.18^*$ & $6.17^*$ & $5.08^*$ & $41.50^*$ & 66.13 & \bf 20.94 & 12.93 \\
\cline{2-9}
& $\Delta_{\text{Shopping - RR}}$ & 43.88   & 45.35   &  5.11   & 37.55     & -45.43 & -3.55 & 48.97 \\
& $\Delta_{\text{Trending - RR}}$  &   9.63   &   11.88   &  -0.89   &-12.47   & -22.01 & 19.53 & 2.48 \\
\bottomrule
\end{tabular}
}
\caption{Prevalence of affiliate links, disclosure rate and compliance across videos in different sources. \dataset{RR} includes both \dataset{Reddit} and \dataset{Random}.
%
%
Bolded cells indicate the highest values for a metric within a comparison.
In each comparison period, all pairwise differences are statistically significant at $p < 0.05$.
A two-sided $z$-test of proportions is used to test the significance of differences in percentage of affiliate videos and channels, as well as the compliance rate across sources. A two-sided Welch's $t$-test is used to test the significance of differences in number of affiliate links and affiliate fraction of links per video.}
\label{tab:rq2:source}
\end{table*}



\para{Built-in platform tools are associated with higher disclosure compliance.}
YouTube’s Shopping program provides a built-in disclosure tool that allows creators to label affiliate products directly through shopping icons, automatically generating a clear disclosure.
Here, we evaluate how such platform-level interventions influence affiliate disclosure compliance.
To assess the impact of built-in disclosure tools, we compare disclosure rates in \dataset{Shopping} against two other sources without built-in disclosure support: \dataset{Trending} (videos promoted by YouTube’s algorithm) and \dataset{RR} (a baseline of randomly selected affiliate videos from \dataset{Reddit}).
\Cref{tab:rq2:source} presents results across two overlapping periods (8/20–12/23 and 1/24–12/24).
We found that \dataset{Trending} exhibited a higher prevalence of affiliate content than other sources but persistently underperformed in disclosure compliance.
Given that Trending videos occupy a prominent position on the YouTube platform, this may indicate a need for the algorithm to better align with regulatory requirements regarding affiliate disclosures
In contrast, \dataset{Shopping} displayed the best compliance among all sources, with a non-compliance rate of only 20.7\% and a clear compliance rate of 61.9\%, suggesting that built-in platform tools can substantially improve disclosure behavior.

\parait{Measuring the effect of Youtube shopping disclosure guidelines and tools.} Building on this observation, we quantitatively assess the effect of YouTube’s built-in disclosure features by computing mean differences in compliance across three metrics.
For each metric m, the effect size is defined as: \( \Delta^{(m)}_{\text{Shopping - RR}} = \mu^{(m)}_{\text{Shopping}} - \mu^{(m)}_{\text{RR}} \), where \( \mu^{(m)}\) represents the  average compliance rate for videos in each source. 
To test statistical significance, we applied stratified sampling by category and channel size, restricted to post-2018 videos, and ran 10K bootstrap iterations to estimate 95\% confidence intervals.
We excluded disclosure guidance as a control variable here because doing so would restrict our analysis only to the top 30 affiliate partners, limiting the generalizability of our findings.
As shown in the bottom rows of \Cref{tab:rq2:bootstrap}, videos in \dataset{Shopping} exhibit much better compliance than those in \dataset{RR}: the non-compliance rate decreased by 44.78\%, while clear compliance increased by 44.65\%, both differences were statistically significant.
These findings suggest a strong association between YouTube’s affiliate disclosure interventions (deployed only for YouTube Shopping) and improved compliance.

\parait{Mechanism behind improved compliance.}  In \dataset{Shopping}, 56.63\% affiliate videos rely solely on the built-in YouTube affiliate tagging tool, without placing any affiliate links in the description. 
While creators can bypass this automatic disclosure by using external affiliate networks and manually adding links, the fact that more than half of affiliate videos used only YouTube’s tool suggests that many creators are willing and able to make clear disclosures when the platform enforces a simple and standardized mechanism.
In addition, 22.11\% of videos use the platform's tools and also include other affiliate links in text description. This hybrid case suggests a clear opportunity: if YouTube is able to expand its disclosure tool to cover all affiliate links, not just those within its own affiliate program, these videos could also achieve clear compliance. 
In other words, platforms like YouTube could substantially increase compliance rate by providing a centralized disclosure approach for all affiliate links. This improvement is much greater than the gains from affiliate partners providing disclosure guidance (+11.11\%) or from regulatory effects observed after 2018 (+9.0\%).

\para{Takeaways.} Overall, our findings reveal that for regulators, although clear compliance rates have more than doubled since 2018 (+9.0\%), the uneven improvements in compliance across content categories and channel sizes, along with persistently vague disclosure language, suggest that their impact alone is limited. 
For affiliate partners, our observational data indicated that the presence of guidance was associated with stronger improvements in compliance (+11.1\%).
Compared to these two groups, YouTube’s own platform-level interventions are more strongly associated with improved transparency and compliance. We observed that the built-in disclosure tools provided by YouTube Shopping are linked to substantially higher rates of clear disclosures (+44.65\%).
While these results do not demonstrate causality, they highlight a strong association between standardized platform-level interventions and improved compliance outcomes. This underscores the potential of scalable, platform-integrated disclosure mechanisms to support more consistent and effective transparency practices.

%% file: related.tex
\section{Related Work}  \label{sec:related}
\para{Platform monetization.}
Early qualitative studies highlighted ethical concerns in influencer-driven marketing. Wu \etal \shortcite{wu_youtube_2016} and James \shortcite{james_real_2017} documented how influencers frequently circumvent disclosure requirements, leveraging YouTube for lucrative yet opaque marketing practices.
Schwemmer \etal \shortcite{schwemmer_social_2018} extended this analysis, raising concerns about the social and economic impacts of product promotions, particularly on younger audiences.
Other studies explored how platform governance related to monetization shapes creator behavior. 
Kumar \etal \shortcite{kumar_algorithmic_2019} and Ormen \etal \shortcite{ormen_institutional_2022} analyzed the 2017 ``Adpocalypse'', showing how sudden policy changes disrupted creator revenue streams. Caplan and Gillespie \shortcite{caplan_tiered_2020} and Dunna \etal \shortcite{dunna_paying_2022} found that monetization policies reflect a tiered governance structure, favoring institutional creators over independent ones.  
More recent large-scale work by Hua \etal \shortcite{hua_characterizing_2022}, Ballard \etal \shortcite{ballard_conspiracy_2022}, and Bertaglia \etal \shortcite{bertaglia2024monetisation} demonstrates a significant shift, finding that creators, particularly those in conspiracy or controversial communities, increasingly rely on alternative monetization, such as affiliate marketing. 
Taken together, this body of work shows how unstable monetization policies have pushed creators toward alternative revenue sources. Building on these insights, our study quantifies the prevalence of affiliate marketing over time and examines how its adoption accelerated after the Adpocalypse.

\para{Disclosures and transparency.}  
Our work was inspired by Mathur \etal~\shortcite{mathur2018endorsements}, who pioneered the empirical analysis of affiliate disclosures on YouTube and found that such disclosures were rare. Building on this, Swart \etal~\shortcite{swart_is_2020} developed an automated tool to detect disclosure statements on the platform.
Cross-platform and cross-language evidence also suggests that under-disclosure is not unique to YouTube. 
In an empirical analysis of Twitter and Instagram, Ershov \etal~\shortcite{ershov2025frontiers, ershov2020effects} found that 96\% of sponsored posts were undisclosed, and as a result, most users failed to recognize them as promotional content.
In the EU, Gui \etal \shortcite{gui2024across} reported low compliance among Dutch influencers on Instagram, YouTube, and TikTok, regardless of language or influencer size.
Similarly, Waltenrath \etal \shortcite{waltenrath2021empirical} showed that disclosure quality on Instagram was generally poor, with Germany achieving only a modestly higher disclosure rate (29\%) compared to around 5\% in other regions.
Several studies have also examined how disclosure transparency affects consumer trust.
The findings are mixed: while some studies suggest that transparent disclosures can enhance credibility and foster positive brand associations \cite{lou2022social, balaban2022role}, other work finds that audiences are wary of influencer marketing, potentially reducing consumer trust and diminishing the credibility of both influencer and brand \cite{van2020effects, cheng2024reputation, vekaria2024inventory}.
We advance this literature by operationalizing the FTC’s disclosure standards and evaluating whether disclosures meet regulatory standards.

\para{Stakeholder interventions.}
A growing body of work emphasizes that disclosure compliance is not solely the responsibility of influencers and argue that platform tools and interface design play a critical role in shaping both the feasibility and quality of disclosures.
Mathur \etal \shortcite{mathur2018endorsements} suggested that built-in platform features, such as Instagram’s paid partnership label, can help facilitate clearer disclosures.
Annabell \etal \shortcite{annabell2024disclosed} conducted a cross-platform study, showing that platform policies and UI design strongly influence disclosure behavior, and argued, using the EU consumer law framework, that effective regulation must take these factors into account.
More recently, Bertaglia \etal \shortcite{bertaglia2025influencer} conducted an empirical analysis of Instagram content posted by EU and US creators and argued for harmonized EU disclosure guidelines, standardized platform templates, and adaptable compliance tools.
Complementing this, an EU Parliament policy report \cite{michaelsen2022impact} recommended compliance APIs for data sharing between platforms and regulators and stressed that responsibility should extend beyond influencers to brands, intermediaries, and platforms.
Extending this line of work, our study presents a large-scale observational analysis of how different stakeholder contexts are associated with disclosure compliance.

%% file: conclusion.tex
\section{Limitations}
\para{Representativeness of our dataset.} 
Although our dataset encompasses 2M videos, it may not completely represent the typical content experienced by YouTube users. Notably, our dataset was over-represented in Gaming, Music, and Entertainment videos, which together account for over 50\% of the total. 
This over-representation stems from two factors: (1) these categories are inherently popular on YouTube\cite{statista2023}; and (2) Reddit, our primary data source, disproportionately surfaced Gaming content, further skewing the distribution.
Furthermore, limited data availability in our \dataset{Shopping} dataset (covering only 2024) and \dataset{Trending} dataset (beginning in 2020) constrained our ability to thoroughly analyze the impact of the "Adpocalypse" and subsequent FTC actions.
However, all analyses presented in this paper are conducted on sufficiently large subsets to support statistically valid conclusions. For instance, we deliberately excluded analyses involving small or fragmented subsets, such as \textit{Trailers}, \textit{Movies}, and \textit{Shows} categories, which each contained fewer than 50 videos.
To assess the robustness of our findings, we repeated all RQ1 analyses 10 times, each using a random 50\% subsample within each subgroup (e.g., video category or channel size). The observed trends remained consistent, with coefficients of variation below 10\%, indicating that the results are not driven by outliers.
For RQ2, our analyses were based on 10k iterations of equal-weight stratified sampling (e.g., balancing by category, channel size, affiliate partner, and video source), ensuring statistically robust and representative estimates.
However, we acknowledge that expanding the dataset to include more heterogeneous content could improve generalizability.

\para{Scope of disclosure analysis.}
We restricted our analysis to textual content, excluding verbal or on-screen visual disclosures due to computational limits and inconsistent quality of caption data.
This focus is justified given that the affiliate links we study appear either in the video description or on YouTube’s Shopping shelf, and FTC guidelines require disclosures to match the format and placement of the promotional content, meaning that a textual disclosure must accompany text-based affiliate link \cite{ftcguide2023updates, FTC_frequently_ask_questions}.  
We also restricted our analysis to English-language text, as our annotation guidelines and classification models are designed for English content.
Additionally, FTC guidelines specify that disclosures should appear “near” the affiliate links and be visible without clicking “more”. Our evaluation addresses the “near” requirement by using 'Explicit' and 'Group' labels to capture whether disclosures are attached to individual or a set of affiliate links. We also use the 'Mixed Group' label when a disclosure does not clearly indicate which links are affiliate links. However, because our classifiers operate on full text descriptions obtained from an automated crawl, we always click on “more” to expand the description section --- an action that users may not perform when they encounter affiliate links. 
Taken together, our analyses represent an \emph{upper bound} on compliance. We do not examine all forms of affiliate promotion, nor do we fully assess the visibility constraints outlined in the FTC checklist.

\para{Limitations of stakeholder influence analyses.} 
While our findings reveal that platform interventions are strongly associated with improved compliance, and we observe statistically significant differences in both platform- and affiliate partner-level analyses, our results are observational and should not be interpreted as causal. Despite our efforts to stratify the sample and control for key confounders (e.g., category, channel size, platform source), unobserved confounders may remain.  
Additionally, because our analysis focuses on the top 30 affiliate partners and post-2018 YouTube videos, the generalizability of these results to the entire platform or to other social media ecosystems remains uncertain.

\section{Discussion}
\para{Implications for efforts to improve disclosure compliance.}
Building on our findings, we advocate for a multi-stakeholder approach centered on platforms, which are best positioned to implement standardized, scalable, and easy-to-use disclosure systems. Regulators and affiliate partners should, in turn, contribute to the design and oversight of these systems. 
To support this model, regulators can support platforms by providing form-like templates and examples of compliant disclosures, which can streamline the design process. Similar strategies have been proposed in both online privacy and influencer marketing contexts, including “privacy nutrition labels” \cite{kelley_2009}, standardized forms \cite{diamantis2023forms}, and standardized disclosure templates \cite{bertaglia2025influencer}.
Further, as suggested by Annabell \etal \shortcite{annabell2024disclosed}, regulators should guide platforms to standardize policy terminology and interface design to ensure consistent compliance.
Affiliate partners should explicitly enforce compliance through their program policies, as our findings show that the presence of guidance is positively associated with higher compliance rates. 
In addition, they could collaborate with platforms by offering compliance-checking APIs, a strategy recommended in an EU policy study for data sharing between platforms and regulators \cite{michaelsen2022impact}, or providing metadata endpoints that help platforms automatically detect affiliate links in undisclosed content.
Finally, to facilitate disclosure monitoring, platforms, regulators, and affiliate partners can integrate automated detection tools, like our textual detection classifier, paired with proportionate enforcement actions and timely/transparent appeal processes.

\para{Key implementation challenges.} For the above collaborative efforts to succeed, several challenges must be addressed. 

\parait{Designing and implementing adequate tools.} While our work is exclusively focused on the text modality, FTC and EU regulations require disclosures across any modality where promotions are present.
Many existing platform disclosure tools remain inadequate \cite{annabell2024disclosed} for this task. For instance, YouTube’s “Includes paid promotion” tag \cite{youtube_endorsements_policy} is an on-video disclosure feature that provides a brief visual cue indicating the video contains promotional content. However, it does not specify which products are being promoted.
Also, per the FTC, when promotions occur within videos, disclosures should preferably appear in both audio and visual formats to reach a wider audience \cite{FTC_frequently_ask_questions}.
In practice, this implies that the disclosure should be visibly and audibly present when a product is being discussed, and removed when the creator transitions to a different product. 
Implementing such fine-grained dynamic disclosures would require detecting product mentions in real time, linking them to affiliate content, and displaying synchronized disclosures, making it a challenging task to automate without significantly increasing creator burden.

\parait{Erosion of trust and engagement.} Prior work shows that disclosures, though intended for transparency, can reduce consumer trust in influencers and brands by undermining perceived authenticity, reputation, and highlighting commercial motives \cite{van2021selling, pfeuffer2020effects, cheng2024reputation}. 
As a result, they warn that creators and platforms themselves may face long-term risks, as declining audience trust and engagement can reduce overall user activity. 
It remains unclear how platforms and creators can balance regulatory compliance with strategies that preserve authenticity and maintain audience engagement.

\parait{Resource limitations.} While multi-stakeholder collaboration could simplify compliance, it would increase stakeholder responsibilities. This challenge is particularly acute given the FTC’s projected reduction in workforce over the next few years \cite{reutersFTC}. However, it is likely that the overall burden is likely to decrease once standards and automated verification tools are established.

\parait{Accuracy of automated compliance detection.} Automated detection of disclosure compliance is likely to be inaccurate in many cases, requiring human moderation to verify accuracy and handle potential issues. This raises classic challenges discussed in the content moderation literature. False negatives allow non-compliance to persist, while false positives can penalize genuinely compliant creators, leading to frustration, reduced trust, or even disengagement \cite{jhaver2019human, hartmann2025lost}. 

\section{Conclusion}
The analyses presented in this paper provide crucial insights into the prevalence of affiliate marketing on YouTube, as well as the rates of compliance with regulatory disclosure requirements.
Unfortunately, they do not show comforting results. Specifically, we find that although the popularity of affiliate marketing is growing on YouTube, the compliance with affiliate-related disclosure standards is alarmingly low. 
More than half of all videos that use affiliate links do not make any disclosures at all. Further, even when disclosures are made, they are rarely in clear compliance with the FTC's guidelines.
These findings highlight the risks posed to consumer protection and trust, especially since the lowest rates of compliance are observed among channels with large (1M+) subscriber bases and on videos actively promoted by YouTube's algorithms.
To address transparency gaps, we evaluated the roles of key stakeholders in improving disclosure compliance. Our results show that YouTube, through platform-integrated disclosure tools, has highly associated with improved compliance. We recommend that regulators and affiliate networks collaborate with the platform to support monitoring and enforcement. Taken together, this paper highlights an urgent need and provides actionable insights to strengthen transparency and safeguard consumer trust in the evolving influencer economy.

%% file: ethics_checklist.tex
\section*{Paper Checklist}

\begin{enumerate}

\item For most authors...
\begin{enumerate}
    \item  Would answering this research question advance science without violating social contracts, such as violating privacy norms, perpetuating unfair profiling, exacerbating the socio-economic divide, or implying disrespect to societies or cultures?
    \answerYes{Yes, see Ethical Considerations section.}
  \item Do your main claims in the abstract and introduction accurately reflect the paper's contributions and scope?
    \answerYes{Yes.}
   \item Do you clarify how the proposed methodological approach is appropriate for the claims made? 
    \answerYes{Yes, see Data and Methods section.}
   \item Do you clarify what are possible artifacts in the data used, given population-specific distributions?
    \answerYes{Yes, we discuss potential artifacts in both the Data and Methods section and the Limitations section.}
  \item Did you describe the limitations of your work?
    \answerYes{Yes, see Limitations section.}
  \item Did you discuss any potential negative societal impacts of your work?
    \answerYes{Yes, see Ethical considerations section.}
      \item Did you discuss any potential misuse of your work?
    \answerNo{Our work identifies and quantifies non-compliance behaviors and does not expose any detection vulnerabilities or evasion strategies. Our findings are intended to support platform governance and regulatory improvements.}
    \item Did you describe steps taken to prevent or mitigate potential negative outcomes of the research, such as data and model documentation, data anonymization, responsible release, access control, and the reproducibility of findings?
    \answerYes{Yes, we describe steps to prevent privacy concerns in the Ethical considerations section. We limit public release any privacy sensitive data. The data collection process and the code for building the tool are detailed in the Data and Methods section to support reproducibility.}
  \item Have you read the ethics review guidelines and ensured that your paper conforms to them?
    \answerYes{Yes.}
\end{enumerate}

\item Additionally, if your study involves hypotheses testing...
\begin{enumerate}
  \item Did you clearly state the assumptions underlying all theoretical results?
    \answerNA{NA}
  \item Have you provided justifications for all theoretical results?
    \answerNA{NA}
  \item Did you discuss competing hypotheses or theories that might challenge or complement your theoretical results?
    \answerNA{NA}
  \item Have you considered alternative mechanisms or explanations that might account for the same outcomes observed in your study?
    \answerNA{NA}
  \item Did you address potential biases or limitations in your theoretical framework?
    \answerNA{NA}
  \item Have you related your theoretical results to the existing literature in social science?
    \answerNA{NA}
  \item Did you discuss the implications of your theoretical results for policy, practice, or further research in the social science domain?
    \answerNA{NA}
\end{enumerate}

\item Additionally, if you are including theoretical proofs...
\begin{enumerate}
  \item Did you state the full set of assumptions of all theoretical results?
    \answerNA{NA}
	\item Did you include complete proofs of all theoretical results?
    \answerNA{NA}
\end{enumerate}

\item Additionally, if you ran machine learning experiments...
\begin{enumerate}
  \item Did you include the code, data, and instructions needed to reproduce the main experimental results (either in the supplemental material or as a URL)?
    \answerYes{We release all code and documentation for building the affiliate link detection tool and disclosure classification models, along with the data used to train the disclosure classifiers in Data and Method section. However, we do not release the training data for the affiliate detection tool, as it includes browser interaction logs that may contain privacy sensitive information.}
  \item Did you specify all the training details (e.g., data splits, hyperparameters, how they were chosen)?
    \answerYes{Yes, see Data and Methods section.}
     \item Did you report error bars (e.g., with respect to the random seed after running experiments multiple times)?
    \answerNo{No, however, we report F1 scores on holdout sets for each classifier. Our evaluation uses a train/test split and 5-fold cross-validation with a fixed random seed for reproducibility.}
    \item Did you include the total amount of compute and the type of resources used (e.g., type of GPUs, internal cluster, or cloud provider)?
    \answerYes{Yes. All models are trained on CPUs with multithreading enabled. No GPU or cloud resources are used.}
     \item Do you justify how the proposed evaluation is sufficient and appropriate to the claims made? 
    \answerYes{Yes, see Data and Methods section.}
     \item Do you discuss what is ``the cost`` of misclassification and fault (in)tolerance?
    \answerYes{Yes, see Data and Methods section.}
  
\end{enumerate}

\item Additionally, if you are using existing assets (e.g., code, data, models) or curating/releasing new assets, \textbf{without compromising anonymity}...
\begin{enumerate}
  \item If your work uses existing assets, did you cite the creators?
    \answerYes{Yes.}
  \item Did you mention the license of the assets?
    \answerYes{Yes. Licenses for open datasets and GitHub repositories, where applicable, are mentioned in the cited sources.}
  \item Did you include any new assets in the supplemental material or as a URL?
    \answerYes{Yes, the URL (\url{https://osf.io/3xpwg/?view_only=fd0cb236942142ad915d9d0c1534a8e0}) is provided in Data and Methods section.}
  \item Did you discuss whether and how consent was obtained from people whose data you're using/curating?
    \answerNo{We work only with publicly available datasets and data we collected ourselves from publicly accessible sources.}
  \item Did you discuss whether the data you are using/curating contains personally identifiable information or offensive content?
    \answerYes{Yes, see Ethical considerations section.}
\item If you are curating or releasing new datasets, did you discuss how you intend to make your datasets FAIR (see \cite{fair})?
\answerNo{We release only the training data for the disclosure classifier and video-level metadata used in our analysis. We document the data and its use in the \textit{Data and Methods} section.}
\item If you are curating or releasing new datasets, did you create a Datasheet for the Dataset (see \citet{gebru2021datasheets})? 
\answerNo{We release only the training data for the disclosure classifier and video-level metadata used in our analysis. We document the data and its use in the \textit{Data and Methods} section.}
\end{enumerate}

\item Additionally, if you used crowdsourcing or conducted research with human subjects, \textbf{without compromising anonymity}...
\begin{enumerate}
  \item Did you include the full text of instructions given to participants and screenshots?
    \answerNA{NA}
  \item Did you describe any potential participant risks, with mentions of Institutional Review Board (IRB) approvals?
    \answerNA{NA}
  \item Did you include the estimated hourly wage paid to participants and the total amount spent on participant compensation?
    \answerNA{NA}
   \item Did you discuss how data is stored, shared, and deidentified?
   \answerNA{NA}
\end{enumerate}

\end{enumerate}

%% file: appendix.tex
\appendix
\renewcommand{\thesection}{Appendix \Alph{section}.}

\section{Affiliate Link Detection Details}
\label{sec:appendix:affiliate_link_detection}

We provide additional details on affiliate link detection, as referenced in ~\Cref{sec:methods:links}.

\para{Interaction graph nodes.}  
The interaction graph contains five node types: network nodes, DOM nodes, decoration nodes, storage nodes, and JavaScript nodes, each representing a trigger or source of a browser interaction. More specifically, network nodes represent the targets of HTTP requests, responses, and redirects; DOM nodes represent elements within the HTML DOM on a webpage (identified by their class ID or element name); decoration nodes represent the keys and values of key-value pairs observed within HTTP requests, responses, or redirects; storage nodes represent data elements stored on the client (browser); and JavaScript nodes represent function calls that occur due to JavaScript execution on a webpage.

\para{Interaction graph edges.}  
We use three types of edges to capture the interactions that occur between different node types.
First, we insert redirect edges between network nodes that are the source and target of a redirect.
Next, we insert modification edges between elements when interactions modify their characteristics—for example, a directed modification edge is added from the node representing {\tt a.com} to its cookie node if {\tt a.com} modifies the cookie values.
Finally, we insert access edges between elements when interactions pass information between them—for example, a directed access edge is added from the node representing a link decorator {\tt key} to a cookie node if information from the cookie is passed into the value for {\tt key}.

\para{Feature extraction.}  
We extract features from each interaction graph in our dataset, focusing on two types: graph structural features and redirect features. Structural features capture the overall structure and connectivity of the interaction graph (e.g., graph density, centrality, average shortest path between all pairs of nodes), while redirect features capture the characteristics of redirects (e.g., number of query parameters attached to network nodes, length of the redirect chain, and Shannon entropy of key-value pairs).

\para{Annotation codebook for ground truth dataset} We manually review the top 1,000 most frequently linked landing domains in YouTube descriptions, along with all domains appearing in their redirect chains. Each interaction graph is labeled as \textit{affiliate} only if at least one URL in the redirect chain meets both of the following criteria; otherwise, the graph is labeled as \textit{non-affiliate}.

\begin{itemize}
    \item \textit{1) Independence Relation.} An affiliate link must direct to a third-party merchant, not the creator’s own merchandise, social media, or co-branded store. To verify this, we manually check whether the YouTuber explicitly states that it is their own merchandise in the video or description, or review the link’s landing page to determine whether its URL or content indicate it is the YouTuber’s own store, a co-branded product page, or a social media profile. If so, we label the interaction graph as \textit{non-affiliate}.
    
    \item \textit{2) Affiliate Identifier.} An affiliate link must contain an affiliate identifier, such as a creator-linked string or a unique token. Following prior work~\cite{mathur2018endorsements}, we analyze frequent domain-parameter pairs, as a high frequency of co-occurrence suggests the parameter may be linked to the affiliate’s identity. We confirm affiliation by checking whether the YouTuber discloses it in the corresponding video. If not, we further verify by examining whether the landing merchant offers an affiliate program.
\end{itemize}

\section{\\Affiliate Disclosure Classification Details}
\label{sec:appendix:affiliate_disclosure_classification}
Below, we provide additional details as referenced in ~\Cref{sec:methods:disclosures}.

\para{Annotating disclosure segments for a ground-truth dataset.} \Cref{tab:disclosure_notation_table} summarizes the annotation codebook used to label disclosure segments. It includes each label, a description with an example, and the associated compliance level for both clarity of compensation and clarity of relationship.

\input{tables/disclosure_notation_table}

\begin{table}[t]
    {
    \small
    \centering
    \begin{tabular}{l c c c} 
        \toprule
        \textbf{Classifier}   & \textbf{Precision}    & \textbf{Recall}     & \textbf{F1}\\
        & \textbf{(\%)} & \textbf{(\%)} & \textbf{(\%)}  \\
         
        \midrule
            Disclosure detection         & 94.7 & 98.9   & 96.8 \\
            \midrule
            (Total) Compensation         & 98.6 & 98.6    & 98.6 \\
            Clear                     & 98.9 & 96.9    & 97.8 \\
            Ambiguous                   & 99.0 & 99.8    & 99.5 \\
            None               & 98.0 & 99.0    & 98.5 \\    
            \midrule
            (Total) Relationship         & 96.9 & 97.0   & 97.0 \\ 
            Explicit            & 99.9  & 98.7    & 99.4 \\    
            Grouped                        & 92.8 & 94.1    & 93.4 \\  
            Mixed group                               & 96.3 & 95.1   & 95.7 \\
        \bottomrule
    \end{tabular}
    }
    \caption{Performance of our three disclosure segment classifiers on their respective holdout validation datasets. For the clarity of compensation and relationship classifiers, we also report the performance for each  label.}
    \label{tab:methods:disclosures:performance}
\end{table}%

\para{Classification result and error analysis.} 
\Cref{tab:methods:disclosures:performance} shows the performance of each classifier on its holdout dataset. We conduct a manual analysis to understand the sources of their misclassifications.
This analysis includes 475 unique segments (187 with disclosures) sampled from the holdout set. We find that the disclosure classifier has zero false negatives and 15 false positives, mainly due to: (1) promotional language with high lexical similarity to true affiliate disclosures (e.g., \textit{“Use my code... to get 10\% off any order”}); and (2) mentions of “affiliate” in non-disclosure contexts (e.g., “Fiverr Affiliate Program”).
For the relationship classifier, most errors occur between the Grouped and Mixed labels, which often differ subtly. Three Mixed cases are mislabeled as Grouped (e.g., \textit{“Note that I may earn a small commission…”}), and one Grouped disclosure is mislabeled as Mixed (\textit{“You should assume we are compensated for any purchases...”}). Additionally, some disclosures are inherently borderline between Explicit and Grouped, such as \textit{“I do get commission when you buy, so thank you”}, leading to ambiguity.
The compensation classifier shows no confusion between Clear and None. All misclassifications involve the Ambiguous class—for example, \textit{“The channel will receive about 1\%”} (Clear mislabeled as Ambiguous) or \textit{“Support-A-Creator code: [ ]”} (Ambiguous mislabeled as None).
These findings suggest that the models are sensitive to explicit monetary cues but struggle with vague or borderline phrasing. Despite these challenges, all classifiers demonstrate high accuracy and are suitable for large-scale compliance analysis.

%% file: tables/disclosure_notation_table.tex
\begin{table*}[t]
\small
{
\centering
\renewcommand{\arraystretch}{1.3}
\begin{tabular}{p{3cm} p{10cm} p{3cm}}
    \toprule[1.2pt]

    \textbf{Label} & \textbf{Description with example} & \textbf{Compliance Level} \\ 
    \midrule
    \multicolumn{3}{l}{\textbf{Clarity of Compensation}} \\ \cmidrule{1-3}
    Clear & Clearly explain that the creator is compensated when a viewer uses the associated affiliate link. Example: “If you click on this link and purchase a product, I get a small commission at no cost to you.” & Compliant \\ 

    Ambiguous & Do not explicitly mention the monetary benefit from viewers’ use of the affiliate link, instead using vague language to imply support. Example: “Support the channel through these links.”   & Possibly compliant  \\

    None & Do no mention of any form of compensation associated with the affiliate links, instead using only technical terms which explicitly disallow by FTC FAQs \cite{FTC_frequently_ask_questions}.  Example is: “This is an affiliate link.”. & Non-compliant \\
    \midrule
    
    \multicolumn{3}{l}{\textbf{Clarity of Relationship}} \\ \cmidrule{1-3}
    Explicit individual & Disclosure placed directly alongside each affiliate link for maximum transparency Example: "This is a sponsored link for {\tt <PRODUCT>}: {\tt <affiliate link>}." & Compliant \\ 

    Grouped & Disclosure involve a single statement disclosing a group of immediately following or preceding affiliate links, allowing readers to easily associate the disclosure with the affiliate links. Example: "I get compensated when you make purchases through the following links: {\tt [list of affiliate links]}."  & Compliant \\ 

    Mixed group & Single disclosure statement for multiple links, including both affiliate and non-affiliate links, which does not allow readers to easily distinguish which links are affiliate links, which may not meet the `conspicuous' standard. Example: "Some of the links in the description are affiliate links." & Possibly compliant \\ 
    \bottomrule

\end{tabular}
}
\caption{Disclosure labels with descriptions, examples, and corresponding FTC compliance levels.}
\label{tab:disclosure_notation_table}
\end{table*}